\documentstyle[prl,aps,multicol,epsfig]{revtex}
\begin{document}
\draft

\title{Fluctuation Dominated Josephson Tunneling with a Scanning Tunneling Microscope}

\author{O.~Naaman, W.~Teizer
and R. C.~Dynes\thanks{Email address: rdynes@ucsd.edu}}
\address{Department of Physics, University of California, San Diego;
9500 Gilman Drive, La Jolla, California 92093-0319}

\maketitle

\begin{abstract}
We demonstrate Josephson tunneling in vacuum tunnel junctions
formed between a superconducting scanning tunneling microscope
tip and a Pb film, for junction resistances in the range 50-300
k$\Omega$. We show that the superconducting phase dynamics is
dominated by thermal fluctuations, and that the Josephson current
appears as a peak centered at small finite voltage. In the
presence of microwave fields (${f=15.0}$ GHz) the peak decreases
in magnitude and shifts to higher voltages with increasing rf
power, in agreement with theory.
\end{abstract}

\begin{multicols}{2}
Scanning tunneling microscopy (STM) has been extensively used in
the study of high-T$_c$ superconductors (HTSC), providing a
spectroscopic tool with unparalleled energy and spatial
resolution. Yet, while superconducting tips have been demonstrated
in the past \cite{Pan1} all STM studies so far have been
performed using normal-metal tips, thus probing only the
single-particle excitation spectrum, the gap structure which is a
consequence of superconductivity, but not the superconducting
(SC) ground state itself. Results from STM measurements of HTSC
show excitation gaps in situations where superconductivity is
believed to be absent (pseudo-gap), such as in vortex cores
\cite{Pan00} and above T$_c$ in underdoped samples
\cite{Renner98}, as well as inhomogenieties in the gap structure
in reportedly high quality BiSrCaCuO crystals \cite{Howland01}.
These results, due to the nature of the measurements, do not
remove the ambiguity with respect to the existence of a finite SC
pair amplitude in the situations studied. In light of this out
standing problem, it seems desirable to have a way to directly
probe the SC pair amplitude with high spatial resolutions on the
order of $\xi$, the coherence length. This can be achieved by
performing STM experiments with SC tips \cite{Balatsky}, measuring
the contribution from Josephson pair tunneling to the total
tunneling current.  In this Letter we report on the observation of
fluctuation-dominated Josephson tunneling in vacuum tunnel
junctions formed between a SC tip and a conventional SC Pb film
at T$\sim$2 K.

\begin{figure}
\epsfxsize=\columnwidth \epsfysize=2.0in \epsfbox{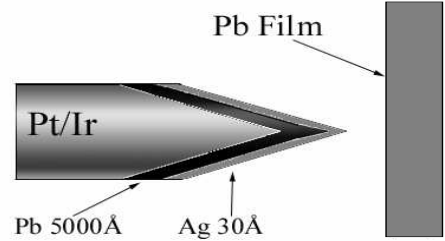}
\caption{Schematic depiction of the experimental setup.}
\label{fig1}
\end{figure}

The present authors have recently developed a method for the
reproducible fabrication of SC STM tips \cite{Naaman01}. Tips
made by deposition of Ag(30 {\AA})/Pb(5000 {\AA}) proximity
bilayer onto conventional Pt$_{0.8}$Ir$_{0.2}$ STM tips exhibit a
well developed BCS gap at low temperatures.  From $\Delta$(T)
measurements, $\Delta_0$ and T$_c$ were estimated to be 1.33 meV
and 6.8 K respectively \cite{Naaman01}. Here we use these tips to
form superconductor/insulator/superconductor (S/I/S) vacuum
tunnel junctions against a Pb film (Fig.\ \ref{fig1}) grown
{\it{in situ}} on a graphite substrate.

\begin{figure}
\epsfxsize=\columnwidth \epsfysize=4.0in \epsfbox{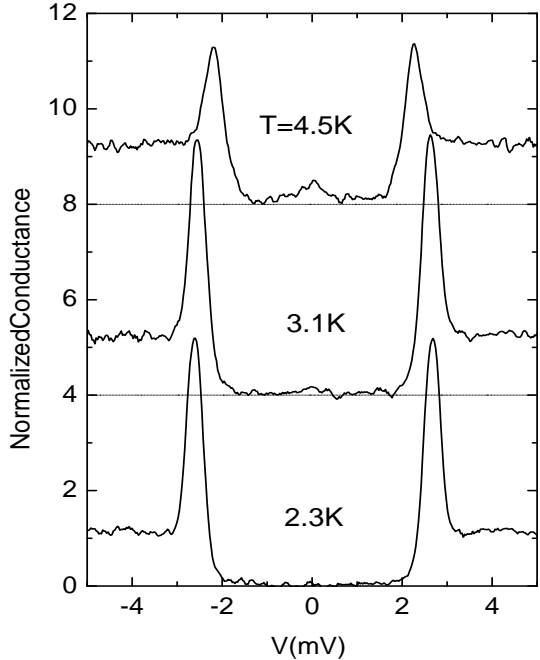}
\caption{Normalized $dI/dV$ curves for a high resistance
($R_N=100$ M$\Omega$) S/I/S junction at various temperatures.
Curves offset for clarity, horizontal lines correspond to zero
conductance of each curve.} \label{fig2}
\end{figure}

For this configuration we observe typical quasiparticle tunneling
conductance spectra for S/I/S junctions with normal state
resistance $R_N=100$ M$\Omega$ as shown in Figure\ \ref{fig2}.
Note the vanishingly small conductance for ${\mid}eV{\mid}<
2\Delta$ at low temperatures, which confirms that tunneling is
the dominant charge transfer mechanism. At higher temperatures,
thermal excitations of quasiparticles across the gap give rise to
a conductance peak at $\pm(\Delta_{Pb}-\Delta_{tip})\approx0$ meV.
At high junction resistances one can expect the SC phases of the
two superconductors comprising the junction to be completely
decoupled, as their coupling, described by the Josephson binding
energy \cite{Ambegaokar63,Pb}
\begin{equation}
E_J=\frac{\pi\hbar}{4e^2}\frac{\Delta(T)}{R_N}\tanh{\frac{\Delta(T)}{2k{_B}T}}
 \label{EJ}
\end{equation}
is vanishingly small compared to $k_BT$. For example, for a
typical junction of 100 M$\Omega$, $E_J/k_B\sim$0.5 mK. Here $T$
is the temperature, $\Delta$ is the SC gap, $R_N$ is the junction
resistance, and all other symbols have their usual meaning.

As the junction resistance is decreased, $E_J$ increases. When
$E_J$ becomes larger than $k_BT$,  a supercurrent of magnitude
$I=I_{c,0}\sin{\phi}$ can flow across the junction at zero
voltage \cite{Solymar}; here $I_{c,0}=2eE_J/\hbar$ and $\phi$ is
the relative phase. When a finite voltage $V$ is applied across
the junction, $\phi$ oscillates according to
$\dot{\phi}=2eV/\hbar$. These are the well known Josephson
equations and the Josephson effect in low resistance junctions
with macroscopically sized electrodes is well understood and well
studied \cite{Solymar}.

The present situation is different- we note that the tip
electrode area is at most 1-2 nm$^2$, thus the very small
junction capacitance $C$ leads to large charging energies of the
junction, $E_C=e^2/2C$. In addition, the junction resistance
cannot be made arbitrarily small and there are two factors
limiting the range of this parameter.  First, because of the
small junction area, the current densities are high and a rough
estimate suggest that junction resistances below few tens of
k$\Omega$ will result in current densities sufficient to destroy
the superconductivity in the tip.  A second limit on the
resistance is the transition from vacuum tunneling to a point
contact regime which was observed to occur around $R_N\sim10$
k$\Omega$ \cite{IBM87}. Experimentally we have been able to
achieve resistances as low as 40 k$\Omega$.

\begin{figure}
\epsfxsize=\columnwidth \epsfysize=4.0in \epsfbox{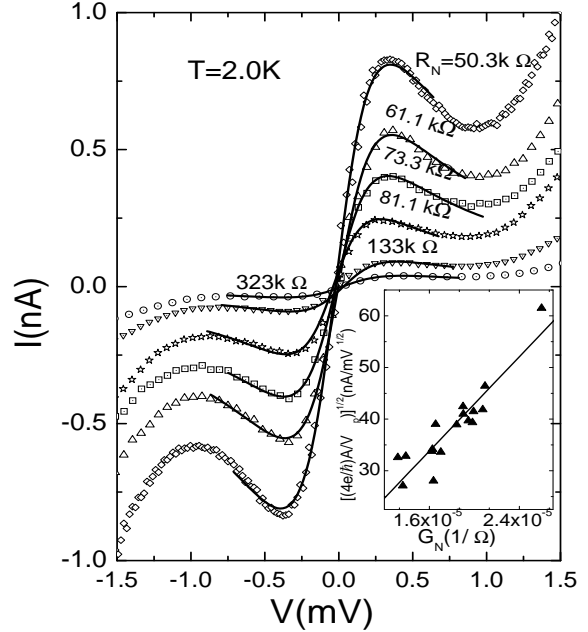}
\caption{Low bias current-voltage characteristics for various
junction resistances at $T=2.0$ K. The data (points skipped for
clarity) is represented by symbols, and the lines represent
two-parameter fits to theory. Inset- $I_{c,0}\times\sqrt{e/k_BT}$
vs. $G_N$ (see text).} \label{fig3}
\end{figure}

Figure\ \ref{fig3} shows typical data from measurements of the
current-voltage characteristics at our base temperature 2.0 K for
various junction resistances.  Note that the data is drawn for
voltages below the sum of the energy gaps on a magnified current
scale. The full $I-V$ curves show a well-defined gap feature at
$\pm(\Delta_{tip}+\Delta_{Pb})=2.6$ meV. The location of the gap
feature does not change with decreasing resistance, and the high
bias part of the curves falls on the same line when the curves
are scaled by $R_N$. The low bias part of these curves, however,
does not reduce to the same line upon scaling. We observe a
current peak centered at finite voltage near zero bias, the
height of which grows when $R_N$ is decreased. Since the SC gap
is temperature dependent, the fact that the gap-edge feature
appears at the same voltage regardless of the junction resistance
is an indication that there is no self heating of the junction
(see also Ref. \cite{heating}).  The current peak observed in
Fig.\ \ref{fig3} cannot, therefore, be attributed to enhanced
quasiparticle thermal excitations across the gap, and must be a
signature of pair tunneling.

One can model the dynamics of the phase $\phi$ in such a loosely
coupled Josephson junction by a point particle moving in a
periodic washboard-like potential landscape
$U(\phi)=E_J(1-\cos{\phi})$ \cite{Likharev86}, and subject to a
stochastic force due to thermal noise
\cite{Ambegaokar69,Ivanchenko69}.  For very high junction
resistances (very shallow potential) the motion of the phase is
completely randomized by thermal fluctuations as $k_BT{\gg}E_J$.
But as the resistance decreases the phase spends on average more
and more time near the minima of $U(\phi)$. We note that even at
the lowest resistance achieved in this work $E_J/k_B\sim1$ K,
thus $E_J$ is comparable to, but still smaller than the thermal
energy in the system. In this case the phase motion can be viewed
as diffusive, and the $I-V$ characteristics of such a junction
have been calculated using this approach by Ivanchenko and
Zil'berman \cite{Ivanchenko69} and Harada {\it{et al.}}
\cite{Harada96} (and also by Ingold {\it{et al.}} \cite{Ingold94}
using a perturbative approach) to have the form
\begin{equation}
I(V)=\frac{I_{c,0}^2Z_{env}}{2}\frac{V}{V^2+V_p^2}, \label{IZ}
\end{equation}
where $V_p=(2e/\hbar)Z_{env}k_BT_n$, considering the thermal
fluctuations as Johnson noise generated by a resistor $Z_{env}$
at temperature $T_n$.  The solid lines in Fig.\ \ref{fig3}
represent fits of our data to Eq.\ (\ref{IZ}) with $V_p$ and
$A{\equiv}I_{c,0}^2Z_{env}/2$ the only fitting parameters. Using
Eqs.\ (\ref{IZ}) and (\ref{EJ}), a plot of
$\sqrt{(4e/\hbar)A/V_p}$ vs.\ $G_N=1/R_N$ (Fig.\ \ref{fig3} inset)
is expected to be linear with zero intercept and slope
$\pi\Delta/2e\sqrt{k_BT_n}$. We obtain from a linear fit
$T_n=5.8\pm0.6$ K, and $Z_{env}=1.5\pm0.2$ k$\Omega$. The value
of $Z_{env}$ is consistent with our experimental setup. The value
of the noise temperature $T_n$ is higher than the actual
temperature of the junction, which is not surprising as the
isolation of the junction from room temperature circuitry is not
perfect. It is also consistent with values reported from similar
measurements on ultra-small, high resistance planar junctions
\cite{Harada96,Danchi84}.

Since our data agrees very well with theory, and it is extremely
difficult to explain our results using a quasiparticle-only
picture, especially once self-heating is ruled out
\cite{heating}, we are led to the conclusion that this is a
fluctuation dominated dc Josephson effect.  This effect was
previously observed in ultra-small planar junctions and is well
documented \cite{Harada96,Buckner70,Muller94}. Furthermore, we
confirm that this effect stems from Josephson tunneling by
measuring the response of the junction in the presence of
microwave fields with frequency $f=15.0$ GHz, fed into a
cylindrical cavity containing the STM by a semi-rigid coaxial
cable antenna \cite{Kleiner96}.

In the absence of fluctuations, when $E_J{\gg}k_BT, E_C$, a
voltage source driven Josephson junction exhibits phase locked
Shapiro current spikes at voltages corresponding to integral
multiples of $\hbar\omega/2e$, where $\omega=2{\pi}f$ is the
angular frequency of the rf field \cite{Solymar,Kleiner96}, and
the height of the k-th order spike $I_k$ at voltage $V_k$ is
proportional to the Bessel function of order k of the reduced ac
voltage induced on the junction.
 When strong thermal
fluctuations dominate the phase dynamics, and especially in our
case where the typical frequency of phase-slip events
$f_{T}{\sim}k_BT_n/h{\sim}1.2\times10^{11}$ Hz is much greater
than the frequency of the applied microwave field
$f=1.5\times10^{10}$ Hz, phase locking cannot be achieved and the
$I-V$ characteristics will exhibit a broad peak instead of
discrete spikes. \begin{figure} \epsfxsize=\columnwidth
\epsfysize=4.5in \epsfbox{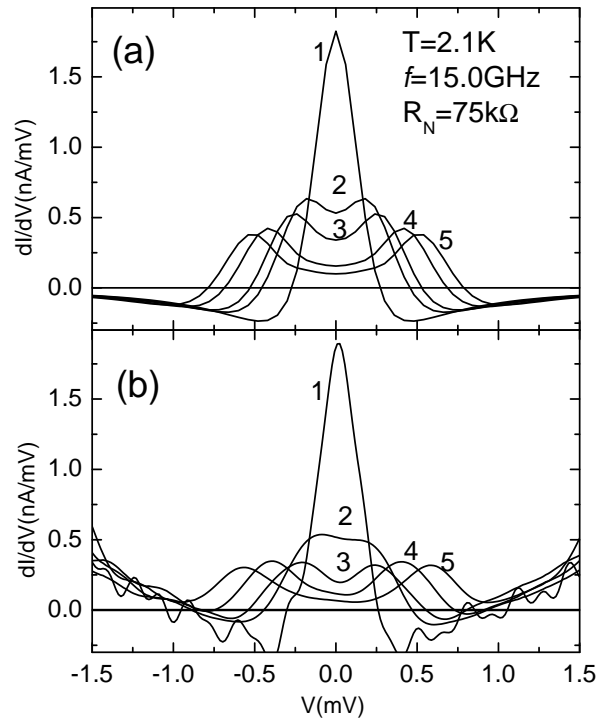} \caption{(a) Calculation
of the conductance at low bias voltages due to Josephson
tunneling in the presence of microwave fields ($f=15.0$ GHz) for
various induced ac voltages on the junction. (b) Numerically
differentiated current-voltage data for a junction with $R_N=75$
k$\Omega$ at $T=2.1$ K. Curves 1-5 correspond to $V_{ac}=0, 0.31,
0.39, 0.54$, and $0.65$ mV respectively.} \label{fig4}
\end{figure} This has been formally calculated by Falci
{\it{et al.}} \cite{Falci91} who obtain the following expression
for the pair tunneling current:
\begin{equation}
I(V)=\sum_{k=-\infty}^{\infty}J_k^2\left(\frac{2eV_{ac}}{\hbar\omega}\right){\times}I_{dc}\left(V-\frac{k\hbar\omega}{2e}\right)
 \label{ac}
\end{equation}
where $J_k$ is the k-th order Bessel function, $V_{ac}$ is the
induced ac voltage on the junction,and $I_{dc}$ is the dc
characteristics of the junction. Numerical evaluation of Eq.\
(\ref{ac}) shows a broad current peak that diminishes in
magnitude and moves away from zero bias as the intensity of the
radiation is increased, and the center of mass of the Shapiro
spikes shifts to higher voltages.  In Figure \ref{fig4} we
present (a) a calculation of the differential conductance $dI/dV$
due to pair tunneling according to Eq.\ (\ref{ac}), and (b) our
numerically differentiated data, which is in very good agreement
with the calculation. Note that the calculation does not account
for the quasiparticle tunneling current background. The observed
conductance peaks shift linearly to higher bias voltages with
increased ac voltage as predicted by Eq.\ (\ref{ac}). At higher
bias voltages (not shown) we observe the expected quasiparticle
tunneling current with a gap edge that smears, due to photon
assisted tunneling, with increasing microwave power.

To summarize, we have demonstrated an STM based Josephson probe.
We have shown that the tunneling characteristics obtained in this
experiment are in good agreement with the model of
fluctuation-dominated Josephson tunneling. While the
fluctuation-dominated dc Josephson effect has been observed
previously in ultra-small planar junctions
\cite{Harada96,Buckner70,Muller94}, this is to our knowledge the
first observation of the corresponding ac effect in the presence
of microwave fields. We expect that a Josephson STM will prove
useful as a probe of the SC pair amplitude on length scales of
the order of the coherence length in high-T$_c$ materials. In
addition, the ease with which the junction resistance in an STM
configuration can be controlled, makes this system favorable for
studying the effects of thermal fluctuations and the Coulomb
blockade (at low enough temperatures) \cite{Likharev85} on the
Josephson phase dynamics.

The authors would like to thank A. D. Truscott, O. Bourgeois, A.
S. Katz, and S. Cybart. Special thanks to S. Nemat-Nasser for
assistance with microwave techniques and equipment. This work was
supported by DOE Grant No. DE-FG03-00ER45853, and by NSF Grant
No. DMR9705180.

\end{multicols}


\begin{references}

\bibitem{Pan1} S. H. Pan, E. W. Hudson, and J. C. Davis, Appl.
Phys. Lett. {\bf 73}, 2992 (1998).

\bibitem{Pan00} S. H. Pan, E. W. Hudson, A. K. Gupta, K.-W. Ng, H. Eisaki, S. Uchida, and J. C. Davis,
Phys. Rev. Lett. {\bf 85}, 1536 (2000).

\bibitem{Renner98} Ch. Renner, B. Revaz, J.-Y. Genoud, K.
Kadowaki, and {\O}. Fischer, Phys. Rev. Lett. {\bf 80}, 149
(1998).

\bibitem{Howland01} C. Howland, P. Fournier, and A. Kapitulnik,
arxiv:cond-mat/0101251 (2001).

\bibitem{Balatsky} J. \u{S}makov, I. Martin, and A. V. Balatsky,
arxiv:cond-mat/0009310 (2000).

\bibitem{Naaman01} O. Naaman, W. Teizer, and R. C. Dynes, Rev. Sci. Instrum. {\bf 72},
1688 (2001).

\bibitem{Ambegaokar63} V. Ambegaokar and A. Baratoff, Phys.
Rev. Lett. {\bf 10} 486 (1963).

\bibitem{Pb} In Pb junctions, $E_J$ is 78.8\% of this value due to strong coupling effects.

\bibitem{Solymar} L. Solymar, {\it Superconductive Tunneling and Applications} (Wiley, New York, 1972).

\bibitem{IBM87} J. K. Gimzewski and R. M\"{o}ller, Phys. Rev. B {\bf 36},
1284 (1987).

\bibitem{heating} S/I/N $dI/dV$ curves taken with the same tip in
the same run over a patch of bare graphite for a wide range of
junction resistances are essentially identical. Since the power
dissipated in the tip due to quasiparticle relaxation is
inversely proportional to the resistance, we can rule out the
importance of self-heating effects.

\bibitem{Likharev86} K. K. Likharev, {\it Dynamics of Josephson Junctions and Circuits} (Gordon and Breach, New York,
1986).

\bibitem{Ambegaokar69} V. Ambegaokar and B. I. Halperin, Phys. Rev. Lett. {\bf 22}, 1364 (1969).

\bibitem{Ivanchenko69} M. Ivanchenko and L. A. Zil'berman, Zh.
Eksp. Teor. Fiz. {\bf 55}, 2395 (1968) [Sov. Phys. JETP {\bf 28},
1272 (1969)].

\bibitem{Harada96} Y. Harada, H. Takayanagi, and A. A. Odintsov,
Phys. Rev. B {\bf 54}, 6608 (1996).

\bibitem{Ingold94} G.-L. Ingold, H. Grabert, and U. Eberhardt,
Phys. Rev. B {\bf 50}, 395 (1994).

\bibitem{Danchi84} W. C. Danchi, J. Bindslev Hansen, M. Octavio,
F. Habbal, and M. Tinkham, Phys. Rev. B {\bf 30}, 2503 (1984).

\bibitem{Buckner70} S. A. Buckner, J. T. Chen, and D. N.
Langenberg, Phys. Rev. Lett. {\bf 25}, 738 (1970).

\bibitem{Muller94} C. J. Muller and R. de Bruyn Ouboter, Physica
B {\bf 194-196}, 1043 (1994).



\bibitem{Kleiner96} R. Kleiner {\it et al.}, Phys. Rev. Lett. {\bf 76},
2161 (1996).

\bibitem{Falci91} G. Falci, V. Bubanja, and G. Sch\"{o}n, Z. Phys. B {\bf 85}, 451 (1991).

\bibitem{Likharev85} K. K. Likharev and A. B. Zorin, J. Low Temp.
Phys. {\bf 59}, 347 (1985).


\end{references}
\end{document}